\documentstyle[pra,aps]{revtex}
\begin{document}
\draft

\title{Large Mass Invariant Asymptotics of the Effective Action}
\author{Alexander A. Osipov\thanks{On leave from the Laboratory 
        of Nuclear Problems, JINR, 141980 Dubna, Russia},
        Brigitte Hiller}
\address{Centro de F\'{\i}sica Te\'{o}rica, Departamento de F\'{\i}sica
         da Universidade de Coimbra, 3004-516 Coimbra, Portugal}
\date{\today}
\maketitle

\begin{abstract}
We study the large mass asymptotics of the Dirac operator with a 
nondegenerate mass matrix $m=\mbox{diag}(m_1,m_2,m_3)$ in the presence 
of scalar and pseudoscalar background fields taking values in the Lie 
algebra of the $U(3)$ group. The corresponding one-loop effective
action is regularized by the Schwinger's proper-time technique. Using
a well-known operator identity, we obtain a series representation
for the heat kernel which differs from the standard proper-time
expansion, if $m_1\ne m_2\ne m_3$. After integrating over the
proper-time we use a new algorithm to resum the series. The invariant 
coefficients which define the asymptotics of the effective action are 
calculated up to the fourth order and compared with the related 
Seeley-DeWitt coefficients for the particular case of a degenerate
mass matrix with $m_1=m_2=m_3$.  
\end{abstract}



The effective action plays the central role in lagrangian quantum 
field theory. Dividing fields into a classical background and quantum
fluctuations and integrating out the last ones one obtains the
effective action which properly accumulates the short distance dynamics
of quantum fields \cite{DeWitt:1965,DeWitt:1967}. In this letter we
study the real part of the one-loop effective action with virtual heavy 
fermions of different masses using the Schwinger's proper-time
technique \cite{Schwinger:1951}. In QCD the nondegenerate mass matrix
of heavy (constituent) quarks results from the spontaneous breakdown
of chiral symmetry, as it is the case, for instance, in the Nambu -- 
Jona-Lasinio model \cite{Nambu:1961}, or from the specially added
invariant term to the QCD Lagrangian that regulates the infrared
behaviour of the effective action \cite{Espriu:1990}. We do not
consider here the manifest chiral symmetry breaking effect on the
effective action \cite{Osipov:2001a}. A careful analysis of this 
problem would lead us too far away from the subject, leaving the 
present result without changes. It might be well to emphasize,
however, that a nondegenerate mass matrix of heavy fermions in
theories with spontaneous breakdown of chiral symmetry appears as a 
consequence of manifest symmetry breaking by the nondegenerate mass 
matrix of light fermions. The mathematical formalism being presented 
in this letter is a necessary element of the approach which faithfully 
mirrors the vacuum structure of such a theory. This approach was
formulated for the theory with an explicit and spontaneous breakdown 
of the global $SU(2)\times SU(2)$ chiral symmetry in
\cite{Osipov:2001a} and generalized to all orders of the asymptotic 
expansion in \cite{Osipov:2001b}. Here we extend this scheme to the
$SU(3)\times SU(3)$ chiral theory where the mass matrix has a form 
$m=\mbox{diag}(m_1,m_2,m_3)$. 

The explicit forms of dominant effective local vertices induced by
virtual heavy fermions in general spontaneously broken gauge theories
have been obtained in \cite{Lee:1982,Lee:1989}. However both the
method of our evaluation of the heat kernel as well as the result,
which is cast as an asymptotic series for the effective action with
the order by order chiral invariant structure of the derived
asymptotic coefficients, are different from the cited papers and new. 
Let us clarify: we do not use the equation of the Schr\"odinger type
for the heat kernel in the first stage of calculations, prefering the 
direct expansion of the heat kernel in powers of background fields,
like, for instance, it has been reviewed in \cite{Ball:1989}. 
Additionally, we suggest a new algorithm for resummations. 
Starting from this place our calculations essentially deviate from 
the known ones. The resummation procedure in \cite{Lee:1982} finally 
leads to an asymptotic series where the n-th term gives the complete 
${\cal O}(m^{-n})$ contributions (including all powers of $\ln m$) to 
the effective action. In chiral theories this structure immediately 
breaks chiral symmetry at each order of $n$. The most direct way to
see this is to consider the theory with a linear realization of chiral 
symmetry. In the spontaneously broken phase the symmetry
transformations of scalar and pseudoscalar fields include mass
dependent parts. Therefore, the asymptotic coefficients must have 
terms with different powers of masses, to be chiral invariant. The 
resummation procedure must take this fact into account properly. In
our approach resummations are organized in such a way that every 
coefficient of the asymptotic series is automatically chiral
invariant, if the Lagrangian possesses this symmetry. We consider this 
property to be a crucial condition on any generalization of the
Schwinger -- DeWitt result. Moreover, the resummation procedure used
in \cite{Lee:1982} is slightly misleading. Indeed, the terms like, for 
example, $[\phi, m^2]$ where $\phi$ is a background field, are
considered to be of $m^2$ order. However, it is clear that the
commutator contributes as the difference of mass squares, i.e., as 
$\sim (m_1^2-m_2^2)$ (in the case of $2\times 2$ mass matrix
$m$). This value can be small and as a result contribute to lower
orders.  

A further approach consists in studying the proper-time asymptotics of 
the heat kernel with arbitrary matrix-valued scalar potentials and 
nontrivial mass matrices as a special case (see, for instance, 
\cite{Gilkey:1998,Ven:1998}) with Seeley-DeWitt coefficients known up 
to $n=5$ for an arbitrary operator of Laplace type. The general
formalism and some references can be found in \cite{Gilkey:1995}, 
chiral gauge theories are reviewed in \cite{Ball:1989}. One should not 
mix the subject of those investigations with the problem considered 
here. We do not study the proper-time asymptotics although in the
limit $m_1=m_2=m_3$ our asymptotic coefficients coincide with
the standard Seeley-DeWitt coefficients. In fact, we use the 
proper-time representation only to separate the field dependent part 
of the heat kernel from the mass dependent one. We do not expand the 
mass dependent piece of the heat kernel in powers of the proper-time; 
instead we manipulate all terms of the proper-time series at
once. This is an essential part of our approach and it means that
there is no way to obtain our result from the well known proper-time 
asymptotics.

After these general remarks, to be more specific, let us consider an 
euclidean quantum field theory with a fermion propagator $D^{-1}$
depending on the background fields collected in $Y$. In the one-loop 
approximation the real part of the corresponding effective action is
given by
\begin{equation}
\label{logdet}  
  W[Y]=-\ln |\det D|=\frac{1}{2}\int^\infty_0\frac{dt}{t}\rho 
       (t,\Lambda^2 )\mbox{Tr}\left(e^{-tD^\dagger D}\right).
\end{equation}
We use here the Schwinger's proper-time representation for the modulus
of the functional fermion determinant \cite{Schwinger:1951}. 
$D^\dagger D$ is a second order elliptic differential operator,
$D^\dagger D=m^2+B$, with $B=-\partial^2+Y$. The symmetry group
acting on background fields is the $SU(3)\times SU(3)$ group of global
chiral transformations. The mass matrix $m=\mbox{diag}(m_1,m_2,m_3)$ 
can be written in the following way 
\begin{equation}
\label{mass}
  m=\sum_{i=1}^3 m_iE_i,\quad (E_i)_{jk}=\delta_{ij}\delta_{ik},
  \quad E_iE_j=\delta_{ij}E_j.
\end{equation}
We find it convenient to use the orthogonal basis, $E_i$, in our 
analysis. The integral over $t$ is divergent at the low limit and
needs to be regularized. This can be done by inserting the regularizing 
kernel $\rho (t,\Lambda^2 )$ with the ultraviolat cutoff parameter
$\Lambda$. In the following we do not need an explicit form for 
$\rho (t,\Lambda^2 )$. Using the technique developed by Fujikawa 
\cite{Fujikawa:1979} one can obtain
\begin{equation}
\label{logdet2}  
     W[Y]=\frac{1}{2}\int d^4x\int\frac{d^4p}{(2\pi )^4}
          \int^\infty_0\frac{dt}{t^3}\rho (t,\Lambda^2)
          e^{-p^2}\mbox{tr}\left(e^{-t(m^2+A)}\right)\cdot 1,
\end{equation}
where $A=B-2ip\partial /\sqrt{t}$. As soon as we have derivatives it 
is necessary to clarify the meaning of the trace in Eq.(\ref{logdet2}).
The coordinate space $\{x\}$ is assumed to have no boundary, so that
one can integrate by parts rendering the functional trace cyclic.
Nevertheless, it is useful to define the cyclically symmetrized trace of 
matrix-valued functions $A_i$ by
\begin{equation}
\label{str}
   \mbox{str}\left(A_1,A_2,\ldots ,A_n\right)\equiv\sum_{perm}\frac{1}{n}
   \mbox{tr}\left(A_1A_2\ldots A_n\right),
\end{equation}
where $perm$ means $n$ possible cyclic permutations inside trace.
For algebraic objects str is equivalent to tr provided the trace is
cyclic. 
 
Since we do not want to expand the mass-dependent part of the heat
kernel in powers of the proper-time, we need the operator identity, which
is well-known in quantum mechanics    
\begin{equation}
\label{factor}
   \mbox{tr}\left(e^{-t(m^2+A)}\right)=\mbox{tr}\left(
   e^{-tm^2}\left[1+\sum^\infty_{n=1}(-1)^n f_n(t,A)\right]\right).
\end{equation}
Here $f_n(t,A)$ is equal to
\begin{equation}
\label{fnA}
   f_n(t,A)=\int^t_0ds_1\int^{s_1}_0ds_2\ldots
   \int^{s_{n-1}}_0ds_n A(s_1)A(s_2)\ldots A(s_n),
\end{equation}
where $A(s)=e^{sm^2}Ae^{-sm^2}$. For our purpose we need the first
five terms of the series. The n-th coefficient is equal to
\begin{equation}
\label{expansion}
   \mbox{tr}\left[e^{-tm^2}f_n(t,A)\right]=\frac{t^n}{n!}
   \sum^3_{i_1,i_2,\ldots ,i_n}c_{i_1i_2\ldots i_n}(t)\mbox{str}
   \left(A_{i_1},A_{i_2},\ldots ,A_{i_n}\right),
\end{equation}
where $A_i\equiv E_iA$, and the totally symmetric coefficients 
$c_{i_1i_2\ldots i_n}(t)$ are equal to
$$
   c_i=e^{-tm^2_i},\quad 
   c_{ij}=\frac{e^{-tm_i^2}-e^{-tm_j^2}}{t\Delta_{ji}},\quad
   c_{ijk}=\frac{2}{t^2}\left(\frac{e^{-tm_i^2}}{\Delta_{ji}\Delta_{ki}}
          +\frac{e^{-tm_j^2}}{\Delta_{kj}\Delta_{ij}}
          +\frac{e^{-tm_k^2}}{\Delta_{ik}\Delta_{jk}}\right),
$$
\begin{equation}
\label{ci}
   c_{ijkl}=\frac{3!}{t^3}\left(
   \frac{e^{-tm_i^2}}{\Delta_{li}\Delta_{ki}\Delta_{ji}}+
   \frac{e^{-tm_j^2}}{\Delta_{ij}\Delta_{lj}\Delta_{kj}}+
   \frac{e^{-tm_k^2}}{\Delta_{jk}\Delta_{ik}\Delta_{lk}}+
   \frac{e^{-tm_l^2}}{\Delta_{kl}\Delta_{jl}\Delta_{il}}\right),
\end{equation}
with the definition $\Delta_{ij}\equiv m_i^2-m_j^2$. It is not 
difficult to write down the expression for any coefficient 
$c_{i_1i_2\ldots i_n}$ in (\ref{expansion}). If the indices are all 
equal, one can obtain that $c_i=c_{ii}=c_{ii\ldots i}$. In the limit 
$m_1=m_2=m_3=M$ this becomes $c_i=\exp (-tM^2)$ and the exponent 
is totally factorized. Thus one can obtain the standard inverse mass 
expansion. In the nondegenerate case for the first five terms in 
(\ref{factor}) we find
\begin{eqnarray}
\label{step1}
   \mbox{tr}\left(e^{-t(m^2+A)}\right)
   &=&\sum^3_{i=1}c_i\mbox{tr}(E_i)-t\sum^3_{i=1}c_i\mbox{tr}(A_i)
      +\frac{t^2}{2!}\left[\sum^3_{i=1}c_i\mbox{tr}(A_i^2)
      +\sum^3_{i\ne j}c_{ij}\mbox{tr}(A_iA_j)\right]
      \nonumber \\
   &-&\frac{t^3}{3!}\left[\sum^3_{i=1}c_i\mbox{tr}(A_i^3)
      +3\sum^3_{i\ne j}c_{iij}\mbox{str}(A_i,A_i,A_j)
      +\sum^3_{i\ne j\ne k}c_{ijk}\mbox{tr}(A_iA_jA_k)\right]
      \nonumber \\
   &+&\frac{t^4}{4!}\left\{\sum^3_{i=1}c_i\mbox{tr}(A_i^4)
      +4\sum^3_{i\ne j}c_{ijjj}\mbox{str}(A_i,A_j,A_j,A_j)
      +\sum^3_{i\ne j}c_{iijj}\mbox{tr}[A_i^2A_j^2+A_iA_j^2A_i
      +(A_iA_j)^2]
      \right.\nonumber \\
   &+&\left.2\sum^3_{i\ne j\ne k}c_{ijkk}\mbox{str}[2(A_i,A_j,A_k,A_k)
      +(A_i,A_k,A_j,A_k)]\right\}+\ldots .
\end{eqnarray} 
Substituting this expression in Eq.(\ref{logdet2}), putting there
$A=Y-\partial^2-2ip\partial /\sqrt{t}$ and integrating over
the four-momenta $p_\mu$
we obtain
\begin{eqnarray}
\label{step2}
   W[Y]&=&\int\frac{d^4x}{32\pi^2}\int^\infty_0\frac{dt}{t^3}
      \rho (t,\Lambda^2)\left\{
      \sum^3_{i=1}c_i\mbox{tr}(E_i)-t\sum^3_{i=1}c_i\mbox{tr}(Y_i)
      +\frac{t^2}{2!}\left[\sum^3_{i=1}c_i\mbox{tr}(Y_i^2)
      +2\sum^3_{i<j}c_{ij}\mbox{tr}(Y_iY_j)\right]\right.
      \nonumber \\
   &-&\frac{t^3}{3!}\left[
      \sum^3_{i=1}c_i\mbox{tr}(Y_i^3-Y_i\partial^2Y_i)
      +\sum^3_{i\ne j}c_{iij}\mbox{tr}(3Y_jY_i^2-Y_j\partial^2Y_i)
      +3\sum^3_{i<j<k}c_{ijk}\mbox{tr}(Y_i\{Y_j,Y_k\})\right]
      \nonumber \\
   &-&\frac{t^3}{12}\left.\left[
      \sum^3_{i=1}c_i\mbox{tr}(Y_i\partial^2Y_i)
      +\sum^3_{i\ne j}c_{ijjj}\mbox{tr}(Y_i\partial^2Y_j)
      \right]\right\}+\ldots .
\end{eqnarray}
Here we have used $Y_i\equiv E_iY$ and considered only terms up to
and including $t^3$ order (we count powers of $t$ as like it would be
in the limit $\Delta_{ij}\rightarrow 0$). Let us remind that the 
effective action is defined up to total derivatives, which can be 
omitted. Let us also stress that the above expression is not a 
proper-time expansion, since the coefficients $c_{i_1i_2\ldots i_n}$ 
are functions of $t$.
 
The integrals over the proper-time $t$ can be simply evaluated and
reduced to combinations of some set of elementary integrals $J_n(m^2_i)$
\begin{equation}
\label{Jn}
   J_n(m^2_i)=\int^\infty_0\frac{dt}{t^{2-n}}e^{-tm^2_i}\rho 
   (t,\Lambda^2),
\end{equation}
where $n$ is integer. However in some sense the result of such 
integrations does not mean much by itself. One has to put the series 
in a form in which every term of the asymptotic expansion is
chiral invariant if $W[Y]$ does. This natural requirement selects from
the infinite number of possibilities for rearrangements in 
Eq.(\ref{step2}) only one as the correct answer.
The main problem here is to find the algorithm which automatically 
yields a chiral invariant grouping for the background fields as well as
the mass dependent factors before them. This problem has been
solved recently \cite{Osipov:2001b} on the basis of the novel 
algorithm for resummations inside the starting expansion (\ref{step2}).
The resummations are determined by the recursion formulas
\begin{equation}
\label{recfor}
   J_l(m_j^2)-J_l(m^2_i)=\sum^\infty_{n=1}\frac{\Delta_{ij}^n}{2^nn!}
   \left[J_{l+n}(m_i^2)-(-1)^nJ_{l+n}(m_j^2)\right].
\end{equation}
In the case under consideration one has to factorize the mass
dependent factors $I_i$, manipulating formula (\ref{recfor}).  
Let us stress that only the following choice for $I_i$ leads to
invariant coefficients $a_i$ in the resulting asymptotic series
\begin{equation}
\label{logdet5} 
   W[Y]=\int\frac{d^4x}{32\pi^2}\sum^\infty_{i=0}I_{i-1}\mbox{tr}(a_i),
   \quad I_i\equiv\frac{1}{3}\sum_{j=1}^3J_i(m^2_j).
\end{equation}
The first four coefficients are equal to
\begin{eqnarray}
\label{coeff}
     a_0&=&1, \quad a_1=-Y, \quad a_2=\frac{Y^2}{2}
         +\frac{\Delta_{12}}{2}\lambda_3Y+\frac{1}{2\sqrt{3}}
         \left(\Delta_{13}+\Delta_{23}\right)\lambda_8Y,
         \nonumber \\
     a_3&=&-\frac{Y^3}{3!}-\frac{1}{12}\Delta_{12}
         \left(\Delta_{31}+\Delta_{32}\right)\lambda_3Y
         +\frac{1}{12\sqrt{3}}\left[\Delta_{13}(\Delta_{21}+\Delta_{23})
         +\Delta_{23}(\Delta_{12}+\Delta_{13})\right]\lambda_8Y
         \nonumber \\
        &+&\frac{1}{4\sqrt{3}}\left(\Delta_{31}+\Delta_{32}\right)
         \lambda_8Y^2+\frac{1}{4}\Delta_{21}\lambda_3Y^2
         - \frac{1}{12}(\partial Y)^2. 
\end{eqnarray}
To obtain this result we used relations between $E_i$ matrices and 
$U(3)$ hermitian generators, $\lambda_0, \lambda_3$ and $\lambda_8$.
One can consider the resulting series an inverse mass expansion,
since $I_{l+1}\sim m_i^{-2l}$ for $l\geq 1$. However one should
remember that the asymptotic coefficients $a_i$ depend on mass
differences. This dependence is completely fixed by the
symmetry requirements. Indeed, it can be verified that if the operator 
$D^\dagger D$ is defined to transform in the adjoint representation 
$\delta (D^\dagger D)=i[\omega,D^\dagger D]$, the coefficient
functions $a_i$ are invariant under the global infinitesimal chiral 
transformations with parameters $\omega =\alpha +\gamma_5\beta$. 
The present result is in agreement with the standard Schwinger-DeWitt 
inverse mass expansion, when $m_1=m_2=m_3$. For the case with 
$\Delta_{ij}\ne 0$ our formula (\ref{logdet5}) is a new asymptotic
series which can be used to construct the low-energy EFT action when 
the local vertices are induced by one-loop diagrams involving heavy
particles with different masses. In the low energy QCD, for instance,
our approach can be used for the Nambu -- Jona-Lasinio type models
with $SU(3)\times SU(3)$ chiral symmetry \cite{Volkov:1984}, or for 
heat kernel one-loop renormalizations in the framework of the chiral
expansion \cite{Espriu:1990}. The difference in the masses of the
nonstrange and strange constituent quarks is large enough, $m_u,
m_d\sim 330$ MeV and $m_s\sim 510$ MeV, so that one may expect 
essential numerical deviations from the results obtained by 
other methods. Finally the method can also be generalized to
accommodate minimal couplings of the loop particles to gauge fields.
Work along these lines is in progress.

This work is supported by grants provided by Funda\c{c}\~{a}o para a 
Ci\^encia e a Tecnologia, PRAXIS/C/FIS/12247/1998, PESO/P/PRO/15127/1999,
POCTI/1999/FIS/35304, CERN/P/FIS/40119/2000 and NATO ``Outreach" 
Cooperation Program.

\baselineskip 12pt plus 2pt minus 2pt

\end{document}